\documentclass
[aps,prl,onecolumn,showpacs,byrevtex,superscriptaddress]{revtex4}

\usepackage{amsmath}

\def\be{\begin{equation}}
\def\eea{\end{eqnarray}}
\def\ee{\end{equation}}
\def\bea{\begin{eqnarray}}
\def\ea{\end{array}}
\def\ba{\begin{array}}

\newcommand{\exval}[1]{\mbox{$\langle \, {#1}\, \rangle$}}
\newcommand{\bel}[1]{\begin{equation}\label{#1}}

\newcommand{\deriv}[1]{\mbox{$\displaystyle\frac{d}{d{#1}}$}}
\newcommand{\pderiv}[1]{\mbox{$\displaystyle\frac{\partial}{\partial{#1}}$}}

\newcommand{\pnderiv}[2]{\mbox{$\displaystyle\frac{\partial^{#1}}{\partial#2^{#1}}$}}

\newcommand{\curl}[1]{\mbox{${\nabla\times} \, {#1}$}}
\newcommand{\divg}[1]{\mbox{${\nabla}\cdot \, {#1}$}}
\newcommand{\grad}[1]{\mbox{${\nabla} \, {#1}$}}

\newcommand{\vel}{{\bf v}}
\newcommand{\vort}{{\vec{\omega}}}

\def\zzz{{\mathchoice {\hbox{$\sf\textstyle Z\kern-0.4em Z$}}
{\hbox{$\sf\scriptstyle Z\kern-0.3em Z$}}
{\hbox{$\sf\scriptscriptstyle Z\kern-0.2em Z$}}
{\hbox{$\sf\textstyle Z\kern-0.4em Z$}}}}

\begin{document}
\title{Spontaneous symmetry breaking and finite time singularities  \\ in $d$-dimensional  
incompressible  flow with fractional dissipation}

\author{G. M. Viswanathan}

\affiliation{Consortium of the Americas for Interdisciplinary Science,
University of New Mexico,
800 Yale Blvd. NE,
Albuquerque, NM 87131, USA}

\affiliation{Instituto de F\'{\i}sica, Universidade Federal de Alagoas,
Macei\'{o}--AL,  CEP 57072-970, Brazil}

\author{T. M. Viswanathan}

\begin{abstract}
We investigate the formation of singularities in the incompressible 
Navier-Stokes equations in $d\geq 2$ dimensions with 
a  fractional Laplacian $|\nabla |^\alpha$.  We derive analytically 
a sufficient but not necessary condition for solutions to remain always 
smooth and
show   that finite time 
singularities cannot form for 
  $\alpha\geq \alpha_c=  1+d/2$. 
Moreover, initial singularities  become 
unstable for $\alpha>\alpha_c$.

\end{abstract}

\revised{\today}
\pacs{
05.70.Jk, %
05.40.Fb, %
47.10.ad, %
}
\maketitle

Scale invariance symmetry~\cite{symmetry,bbm,bunde-havlin}
holds approximately for the nonlinear Navier-Stokes equations for
incompressible flow of a Newtonian fluid~\cite{ns1,ns2}.  The
nonlinearity and the scale invariance together can create conditions
for the energy to cascade down to increasingly finer spatial and
temporal scales, e.g., turbulence~\cite{ns2}.
In two dimensions, singularities cannot form~\cite{ns1}.  
However, more
than a century since the discovery of these nonlinear parabolic
partial differential equations, the question remains unanswered 
whether or not
singularities can form in three 
dimensions, due to the crucial role
played by scale invariance.  Such  fundamental problems remain the
subject of ongoing investigations, due to their importance to a number
of fields of physics 
and 
mathematics~\cite{ns1,ns2,phys-sis,phys1,phys2,math1,math2,math3,math4,math5,gen-lapl1,ms,bkm,sing1,sing2}.

Here we address the general question of under which conditions
dissipation can  overcome inertial effects to prevent the
formation of singularities in finite time.  
We answer this question by
 generalizing the problem via
a fractional
Laplacian operator,  and then
analytically deriving  hard inequalities based on the 
fact 
that no singularity 
can form  provided all  partial space derivatives  of all orders of the velocity field remain finite for all time.
In terms of the  
$d$-dimensional
Fourier transform ${\bf \tilde v(k,}t)$ of the velocity field ${\bf v(x,}t)$, we can write 
\bea 
&~& \!\!\! \!\!\!\! \!\!  \!\!
(2\pi)^{d/2}
\bigg|   \pnderiv{n_1}{x_1} 
\pnderiv{n_2}{x_2} \ldots 
\pnderiv{n_d}{x_d} 
{\bf v}({\bf x})~\bigg| \nonumber   \\ &=&
\bigg|\int _{-\infty}^\infty{\bf d}^d {\bf k}  ~ 
\exp[{\mathtt i} {\bf k}\cdot {\bf x} ] {\bf \tilde v}({\bf k}) 
({\mathtt i}k_1)^{n_1}
({\mathtt i}k_2)^{n_2} \ldots
({\mathtt i}k_d)^{n_d} \nonumber \bigg| \\
&\leq&
\int  _{-\infty}^\infty{\bf d}^d {\bf k}  ~ 
{k^{n_1+n_2+\ldots n_d}}
~ |{\bf \tilde v}({\bf k})|
,
\quad k=|\bf k| 
\; \nonumber .
\eea
 To simplify the notation, let
$\exval{A,B}$ represent the absolute integral  $\int _{-\infty}^\infty |A({\bf k})|  |B({\bf k})| ~{\bf d}^d {\bf k}$. 
Then, the above inequality becomes 
\be
\bigg|\pnderiv{n_1}{x_1} 
\pnderiv{n_2}{x_2} \ldots 
\pnderiv{n_d}{x_d} 
{\bf v}({\bf x})\bigg| 
\leq
\exval{k^{n} , \bf  \tilde v},\quad n=\sum_i^d n_i  \;\;.
\label{eq-1}
\ee
The integrated quantities represent 
modified 
 Fourier analogs of the 
statistical moments typically encountered in the study of anomalous diffusion and L\'evy flights~\cite{bachelier,levy,metzler,kenkre,mfs,sokolov,barkai,nature,alz}. 
 Indeed,  our  inspiration comes from the theory of random walks.
In what follows, we 
use this approach to study the formation of singularities, 
by deriving  bounds on the rate of growth of these 
``absolute moments'' $\exval{k^n,\bf \tilde  v}$.
From Eq. (\ref{eq-1}), we can conclude that 
no singularity can form provided all  such absolute moments in Fourier space 
 remain finite.

The dissipation term in 
the Navier-Stokes equations 
contains a Laplacian for  incompressible Newtonian fluids.  Generalizing the Laplacian operator  $\nabla^2$  via Riesz fractional 
derivatives~\cite{metzler}, we obtain  
equations for anomalous dissipation~\cite{gen-lapl1}:
\bea
\rho\bigg[ \pderiv{t} v_i({\bf x},t) + {\bf v} \cdot {\bf \nabla} {v_i({\bf x},t)} \bigg]
&=& \mu_\alpha |\nabla |^\alpha v_i({\bf x},t) \nonumber   + f_i({\bf x})  \\
& & 
- \pderiv{x_i} p({\bf x},t)  \label{eq-ns}
 \\
\divg{{\bf v}} ({\bf x},t) &=& 0\;\; , \label{eq-div}
\eea
where $p$ denotes pressure, 
$\mu_\alpha$ the viscosity, $f$ the external forces, $\rho$ the density,
$\alpha>1$ the order of the fractional Laplacian and $i=1 \ldots d$.
Fractional dissipation can arise 
 due to  
anomalous (i.e., non-Fickian) transport of momentum, e.g., 
in non-Newtonian fluids.
Substitution  into Eq. (\ref{eq-ns}) 
of 
the scale transformation $\bf v\rightarrow \lambda v$,  ${\bf x} \rightarrow 
\lambda^{\frac{1}{1-\alpha}} \bf x$, $p\rightarrow \lambda^2 p$, and 
$t\rightarrow \lambda^{\frac{\alpha}{1-\alpha}} t $, by a factor $\lambda$, leaves the
Navier-Stokes equations unchanged (except for $f_i$).  These
scale free properties leave open the possibility that solutions may
conceivably possess structure at arbitrarily small scales, such that
for $\alpha=2$, $d \geq 3$ we still do not know whether or not singularities can
form in finite time
given smooth
initial conditions and external forces~\cite{ns2,math1,gen-lapl1,ms}.  
Our strategy  consists  in looking for  a spontaneously broken 
scale invariance symmetry.

Fourier transforming these equations to eliminate the space
derivatives, we obtain 
\bea \rho \bigg[ \pderiv{t} \tilde{v}_i({\bf
k},t) &+& \sum_j^d \tilde{v}_j({\bf k},t ) * \big({\mathtt i}k_j \tilde{v}_i({\bf
k},t)\big) \bigg] \nonumber \\ \quad \quad &=& - \mu_\alpha k^\alpha
\tilde{v}_i({\bf k},t)
 - {\mathtt i}k_i \tilde p({\bf k},t) + \tilde f_i({\bf k}) \nonumber \\ 
  \label{eq-fourier} ~
  \\ \quad {\bf k}\cdot {\bf \tilde{v}}({\bf k},t) &=& 0 \;\;, 
  \eea
	with the nonlinearity becoming a nonlocal convolution term in
	Fourier space.  
	Here, \mbox{${ \mathtt  i}=\sqrt{-1}$} whereas $i$ denotes a component index.
	We now note that for any function $g(t)$, if 
	\mbox{${d}g/{d}t = A(t) - B(t) g(t)$} and $B\geq0$ then \mbox{$
	\frac{d}{dt}|g| \leq |A| -B |g |$.}    
Identifying   $g$ with $\rho \tilde v_i$
and
	$B$ with $\mu_\alpha k^\alpha$, we get from 
Eq. (\ref{eq-fourier}),
\bea \rho \pderiv{t} |\tilde{v}_i({\bf k},t)| &\leq& \rho\bigg|\sum_j
\tilde{v}_j({\bf k},t ) * \big({\mathtt i}k_j \tilde{v}_i({\bf k},t)\big) \bigg|
\nonumber \\ &~& - \mu_\alpha k^\alpha |\tilde{v}_i({\bf k},t)| \\
\nonumber &~& + | k_i \tilde p({\bf k},t)| + |\tilde f_i({\bf k}, t)|
\eea
Multiplying by $k^n$ ($n=0,1,2,\ldots$)
 and integrating out $\bf k$ over the entire
Fourier domain, we get,
\bea 
& & \rho\deriv{t} 
\int_{-\infty}^\infty
{\bf d}^d {\bf k}~ k^n |\tilde{v}_i({\bf k},t)| \nonumber \\ 
&~&
\leq \rho \int_{-\infty}^\infty {\bf d}^d {\bf k} ~ k^{n} \sum_j
\bigg|\tilde{v}_j({\bf k},t )*\big(k_j \tilde{v}_i({\bf k},t)\big)
\bigg| \nonumber \\ &~& \quad - \mu_\alpha\int_{-\infty}^\infty {\bf d}^d
{\bf k}~ k^{\alpha+n} |\tilde{v}_i({\bf k},t)| \nonumber \\ &~&
\quad + \int_{-\infty}^\infty {\bf d}^d {\bf k} ~ k^n \big(
|k \tilde p_i({\bf k},t)| +|\tilde f_i({\bf k},t)|\big) \label{moment-long} \eea
Using the shorthand notation $\exval{\cdot,\cdot}$ introduced earlier,
Eq. (\ref{moment-long})   leads to,
\bea
\rho \deriv{t} \exval{k^n, \tilde{v}_i}   &\leq&    
\rho \sum_j  
\int {\bf d}^d {\bf k}'
\exval{|{\bf k}+{\bf k'}|^{n},{\tilde{v}_j}} 
 \big |k_j' \tilde v_i({\bf k'},t) \big |  \nonumber \\
& & 
 -  
\mu_\alpha  \exval {k^{\alpha+n},{\tilde{v}_i}} +C_{i,n}(t)~  \nonumber \\
  &\leq&
\rho  \sum_j  
  \int {\bf d}^d {\bf k}'
     \sum_\ell^n  {n  \choose  \ell} \bigg[ \nonumber \\ & & 
\exval{k^\ell,\tilde{v}_j} 
 |k'|^{n-\ell}
\big | k_j'  \tilde{v}_i(  {\bf k'},t)    \big | \bigg]   
\nonumber \\
& & -  
\mu_\alpha \exval {k^{\alpha+n},{\tilde{v}_i}} + C_{i,n}(t)~   \nonumber
\eea
where $
 C_{i,n}(t) \equiv  \exval {k^{n+1}, \tilde p_i} + \exval{k^n,\tilde f_i} 
$.
We thus arrive at the key inequality 
from 
which we will  derive three interesting results:
\bea
\rho \deriv{t} \exval{k^n, \tilde{v}_i}   &\leq&    
\rho \sum_j  
     \sum_\ell^n  {n \choose \ell}
\exval{
k^\ell, \tilde{v}_j}  \exval{k^{n-\ell+1}, \tilde{v}_i}    \nonumber \\ & &  -
\mu_\alpha  \exval {k^{\alpha+n},{\tilde{v}_i}} + C_{i,n}(t) \;\;\label{moments} .
  \eea
No approximation has gone into the exact derivation of 
this hard inequality.
Given smooth initial conditions, 
the following picture thus emerges of the cascade process: excitations
of the Fourier modes can pseudorandomly ``diffuse'' anomalously from
near the origin $k=0$ outwards, towards large $k$.  We now ask whether
for bounded energy $E$ the moments $\exval{k^n,\tilde{v}_i}$ can
diverge in finite time.  They cannot diverge 
at infinite time because nonzero dissipation implies that energy
decays to zero.
 
The terms 
$C(t)$ do not matter 
in the context of singularities. We briefly review
the known reasons for their irrelevance.
Due to lack of a separate pressure evolution equation and  
the special role of
pressure in incompressible flow,
 we know that the absolute moments of pressure cannot
diverge unless the moments of velocity diverge.  
Indeed, Leray projection methods 
  can eliminate pressure~\cite{math1,ns2}, because 
in incompressible flow pressure 
merely serves to ensure Eq. (\ref{eq-div}).
It does not contribute to the cascade process,
e.g., in $d=3$ pressure does not appear in the vorticity equation (see
below).
Similarly, we can ignore smooth external force terms because they do not 
possess scale invariance.  By their very
definition, smooth 
 external forces have a minimum scale,
i.e., they  act only within a well defined band of spatial
frequencies, so $\exval{k^n,f_i}<\infty$ for all $n\geq 0$.  From now
on, we altogether ignore the terms $C(t)$.

Indeed, the interesting physics only concerns which process wins the
 competition between the dissipation effect that damps high frequency
 (large $k$) modes and the nonlocal interaction that transfers (or
 cascades) the energy to the higher frequency modes.  In Inequality
 (\ref{moments}), the $n$-th moment \exval{k^n, \tilde{v}_i} can only
 grow provided  the dissipation term containing $\alpha$ remains
 smaller than the inertial terms containing products of moments.   
 We thus arrive at  a sufficient but not necessary condition for 
 all moments to remain finite:
 \be
 \lim_{\footnotesize ~~~\exval{ k^n,\bf \tilde v}\rightarrow \infty}
 \frac
 {\exval{k^\ell,\tilde v_j}\exval{k^{n-\ell+1},\tilde v_i}}
 {\exval{k^{n+\alpha},\tilde v_i}}
= 0\;\;,
\label{result1}
\ee 
for all $n, i, j$ and $0\leq \ell \leq n.$
 This condition guarantees that structures of arbitrarily small spatial scale
 will dissipate.
 However, 
  if the condition
does not hold, then  
  Inequality (\ref{moments})  may in fact allow  a finite
 time singularity of 
  the form $\exval{k^n, \bf \tilde{v}} \sim
 |t_c-t|^{-\delta_n}$.
The
original question of whether or not singularities can form now reduces
to the more manageable problem of whether or not  
Eq. (\ref{result1}) holds.

We next   look for  evidence of spontaneous
symmetry breaking.
As the excitations in the Fourier modes cascade to larger $k$, the
moments grow due to two 
related but 
distinct reasons: (i) an
``entropic''   effect, due to the pseudorandom or chaotic
spreading or nonlocal diffusion  of the excitations in Fourier space and (ii) an energy
renormalization effect, due to conservation of energy, which
leads to approximate $L^2$ normalization of $\bf \tilde v(k)$:
 $\exval{\bf \tilde v,\tilde v} \leq E/\rho$.  The equality holds 
only at $t=0$ 
 due to energy dissipation.
As an illustrative example of effect (ii), 
if the energy in a single mode with initial amplitude $\tilde{v}_0$
 becomes evenly split into $N^d$ modes, then the sum of the amplitudes
 of the $N^d$ modes grows to $N^{d/2}\tilde{v}_0$.  
 Moreover,
 this energetic growth affects 
not  
 only $\exval{k^0,\bf \tilde v}$ but all moments 
 $\exval{k^n,\bf
\tilde  v}$.
  The interplay between these entropic and energetic effects lies at
the root of the the fundamental physical mechanism of singularity
formation. 
 As
 the moments grow sufficiently 
 large, we know that \mbox{$\exval{k^m, \bf \tilde{v}} \gg
 \exval{k^n, \bf \tilde{v}}$} if $m>n$.   
 The
 highest moment appears in the dissipation term, since $\alpha>1$.
Nevertheless, overgrowth of 
the numerator of Eq. (\ref{result1})  relative to
the higher moment of order $n+\alpha$ in the denominator
may cause the limit not to vanish.
We may then  lose all control and a finite time singularity becomes
a possibility.
  
 The worst case scenario of fastest possible overgrowth for the lower
moments in 
the numerator of Eq. (\ref{result1}) 
corresponds 
to isotropic cascade with a moment
$\exval{k^m,\bf \tilde{v}}$ that diverges in finite time.
Moments can diverge only due to fat tails 
(i.e., asymptotic power law decays) 
in $|\bf \tilde v(k)|$.
Hence,  we arrive at 
 a  multifractal scaling relation for moments
in the limit of
large $\exval{k^m,\tilde{v}_i}$ for the worst case scenario:
\be 
\exval{k^n,\tilde{v}_i} \sim
\exval {k^m,\tilde{v}_j}^{\mbox{$\frac{n+d/2}{m+d/2} $} },
\quad n\geq m\;\;,
 \label{mf}
 \ee 
 for all $i, j$ and $m\geq 0.$
 This scaling relation follows 
 directly 
  from 
 conservation of energy
and 
the terms  $d/2$ in the exponent 
come from the energy renormalization effect, as
discussed earlier.   Substituting Eq. (\ref{mf})  into
(\ref{result1}), we obtain the 
value of the 
upper critical $\alpha_c$ that guarantees finite moments: 
\be
\alpha_c = 1+d/2 \;\; .\label{critical} \ee 
This critical value  constitutes our second important result: singularities
cannot form if $\alpha \geq \alpha_c$. 
 For the marginal
case $\alpha= \alpha_c$ we 
can see from Inequality (\ref{moments})
that 
dissipation
will hold off the singularity by ``buying time'' for the energy to
decay.  In fact, the total energy 
decays always with time, as mentioned earlier.
However,
we ignore this effect except for the marginal case because we can
easily show its relative impotence to prevent singularity formation in
finite time for the non-marginal cases.  
We thus obtain 
$\alpha_c=5/2$ for the 
important 
special case $d=3$, in
agreement with the result obtained in refs.~\cite{gen-lapl1,ms} using
different approaches.   Eq. (\ref {critical}) also correctly gives  the
 known
value $\alpha_c=2$ in $d=2$.
Moreover,  
Eqs. (\ref{result1}) and (\ref{mf}) together 
 guarantee 
  that 
pre-existing singularities become unstable
under small perturbations
 for 
$\alpha>\alpha_c$.

We next proceed to our third  and final  result. 
Critical points often (but not always)
separate phases with different symmetries.
Intuitively,  we 
expect that a spontaneously broken scale invariance symmetry 
should  lead to the emergence of a 
special 
characteristic scale that breaks the symmetry.
We next show that this indeed happens.
If $\alpha\neq\alpha_{\mbox{\tiny c}}$ then
Inequality (\ref{moments}) 
shows the
existence
of a characteristic length 
scale $\eta_c=2\pi/k_c$ for which the inertial
and dissipative effects can cancel each other.
Since only an  inequality relation holds, rather than 
 equality relation,
therefore we cannot 
make an  estimate of 
this characteristic scale.
For $\alpha<\alpha_c$ ($\alpha>\alpha_c$), we can
 make
the adimensional equivalent  of 
$\mu_\alpha$ very large (small) to obtain
the dependence of $\eta_c$ on known quantities:
\bea  
\eta_c   &\propto&
\left(\frac{E\rho}{\mu_\alpha^2}\right)^{\frac{1}{2(\alpha_c-\alpha)}} ,
\quad \eta_c > 0
\;\;
.
\label{eq-third}
\eea
Note that  the units of $\mu_\alpha$ depend not only on $d$ 
but also on $\alpha$.
For $\alpha>\alpha_c$,  the length $\eta_c$ represents the smallest
scale to which energy can cascade.
For $\alpha<\alpha_c$, the value  $\eta_c$
represents
the scale  below which inertial effects may dominate,
conceivably 
allowing  singularities.
For $d=3$ and $\alpha=2$, 
in the limit of 
low Reynolds number $Re$,
we get 
$\eta_c \sim Re^2$.  No matter how viscous or how small we take $Re$,
  once a structure
forms at length scales smaller than $\eta_c$, we cannot rule out further
cascade to ever finer scales.  This somewhat counter-intuitive result
suggests that although the parabolic 
Navier-Stokes equations represent a
singular perturbation of the hyperbolic 
Euler equations, yet  for certain classes of
solutions
the behavior of ultra-fine structures may 
possibly 
remain
relatively unperturbed by dissipation at length scales $\eta \ll \eta_c$ for $\alpha<\alpha_c$.

We briefly discuss each of these three results. 
Note that our first result, Eq. (\ref{result1}), 
 represents a sufficient 
but not  necessary 
condition for solutions to remain free of singularities.
We emphasize 
the  impossibility of  
 obtaining  the analogous 
  necessary condition from Inequality (\ref{moments}).
  The second result,
concerning the upper critical dimension, shows precisely the crucial
role that dimensionality has on singularity formation. 
This advance has deep physical significance because of the
importance of $\alpha_c$ in relation to the Hausdorff dimension 
of the
singular set at the time of first blow up~\cite{gen-lapl1}.  The third result  
has a bearing on spontaneous symmetry breaking, which we address 
next.

Although
the Navier-Stokes equations 
possess scale invariance symmetry,
 yet
our results, (\ref{critical}) and (\ref{eq-third}),
indicate  spontaneous symmetry breaking.
A trivial characteristic length 
scale $\eta=0$ always 
exists for any $\alpha$, corresponding to singular solutions, 
but our results indicate the existence of a second, nontrivial,
 length scale $\eta_c$.
Whereas Inequality (\ref{moments}) does not rule out 
stable singularities for $\alpha<\alpha_c$, on the other
hand for $\alpha>\alpha_c$ dissipation always overcomes
inertial effects at scales smaller than $\eta_c$, 
rendering
singularities unstable.   
The characteristic scale $\eta_c$
spontaneously 
breaks scale invariance symmetry. 
In contrast, 
for  $\alpha=\alpha_c$,
inertia and dissipation have exactly the same scaling, so that 
the system becomes truly 
scale invariant, i.e., the original scale invariance
symmetry remains unbroken.

In this context, we note that 
  T. Tao \footnote{http://terrytao.wordpress.com  .} has 
   pointed out the crucial role of energy
   and 
    how it  behaves 
   under scale transformations.
If the velocity in a region
could evolve to become 
larger by a factor $\lambda$, as discussed earlier,
then a (possibly intermittent) cascade of ever smaller
rescaled solutions could repeat the process iteratively 
 until the velocity blew
up in finite time.
For $\alpha=2$ and $d=2$, the energy remains invariant under
scale transformations, hence energy considerations
by themselves 
 prohibit 
the formation of singularities.
 On the other hand, in $d=3$  the energy goes to zero as
 $\lambda\rightarrow \infty$, so energy arguments fail.
In  the general case of Eq. (\ref{eq-ns}),
 the energy 
transforms according to $E\rightarrow \lambda^{2-\frac d{(\alpha-1)}}E$,
which  we can express in terms of $\alpha_c$ 
 as  $E\rightarrow \lambda^{\frac{2(\alpha-\alpha_c)}{\alpha-1}}E$.
Thus,  the physical interpretation
 of 
 $\alpha_c$   becomes clear:
for  $\alpha\geq \alpha_c$ 
singularities  become energetically forbidden.

Finally, we comment on  the 
case $\alpha=2$,   in terms of 
the relevant dynamical 
processes.
  In $d=3$, the vorticity
$\vort \equiv \curl{\bf v}$ satisfies the vorticity equation \be
\pderiv{t} \vort + \vel \cdot \grad{\vort} = \vort\cdot \grad{\vel}
\;\;, \ee for inviscid flow~\cite{ns2}.  
The last term generates vorticity
stretching, which can increase vorticity and excite higher frequency
 modes.
In $d=2$ the vorticity stretching term vanishes, but in $d=3$ it
allows energy and vorticity to  cascade and this behavior carries over
to viscous flow. 
 The well known
criterion of Beale-Kato-Majda~\cite{bkm} states that singularities in
the Euler equation in $d=3$ can only form if the time integral of the
maximum vorticity diverges. 
Whether or 
not singularities can form in $d=3$ for Euler flow, however,
still 
remains an open question. 
 In $d>3$ we expect even more violent energy
cascades.  So  Eqs. (\ref{critical}) and  (\ref{eq-third}) suggest
 that in $d=3$  perhaps the more important question concerns singularity formation for the Euler equations, since $\alpha<\alpha_c$.
Eq. (\ref{eq-third})  conceivably makes room for singularities in Euler
flow, if they in fact occur,  to persist in the dissipative case.  
Numerical results have suggested (but not shown)  that the Euler
equations in $d=3$ may in fact allow singularities~\cite{sing1}.
Singularities 
possibly 
persist even in the Navier-Stokes case~\cite{sing2} 
under special
circumstances.

In summary, we have investigated the behavior of solutions of
Navier-Stokes equations 
with fractional Laplacians $|\nabla|^\alpha$
in $d$ dimensions  and
shown 
analytically 
that singularities cannot form for  $\alpha\geq\alpha_c$.
Our approach relies on studying 
 the growth of absolute moments $\exval{k^n,\bf v}$
in Fourier space. 
We
have used neither mean field approaches nor introduced
stochasticity. Our results represent hard limits, since they
correspond to the worst imaginable scenarios.
We  hope that  the  approach
developed here 
may inspire  future application towards 
a deeper
understanding of other nonlinear partial differential
equations.

We are very grateful to VM Kenkre for help at all stages
and thank the Consortium of the Americas for Interdisciplinary Science
for its hospitality.
We thank 
G Buendia,
IM Gleria,
HD Jennings,
MGE da Luz 
for discussions 
and 
CNPq (Process 201809/2007-9) and  
NSF (Grant no. INT-0336343) 
 for financial support.

\end{document}